\documentclass[%
 reprint,
 superscriptaddress,
 amsmath,amssymb,
 aps,
 floatfix,
]{revtex4-2}

\usepackage{graphicx}
\usepackage{dcolumn}
\usepackage{bm}
\usepackage{subfig}
\usepackage{float}
\usepackage{caption}
\usepackage{siunitx}

\usepackage{xcolor}

\begin{document}

\preprint{APS/123-QED}

\title{{Polarization-entangled photons from a whispering gallery resonator}}

\author{Sheng-Hsuan Huang}
 \email{sheng-hsuan.huang@mpl.mpg.de}
\affiliation{Max Planck Institute for the Science of Light, Staudtstrasse 2, 91058 Erlangen, Germany}
\affiliation{Department of Physics, Friedrich-Alexander-Universität Erlangen-Nürnberg, Staudtstrasse 7/B2, 91058 Erlangen, Germany} 
\author{Thomas Dirmeier}
\affiliation{Max Planck Institute for the Science of Light, Staudtstrasse 2, 91058 Erlangen, Germany}
\affiliation{Department of Physics, Friedrich-Alexander-Universität Erlangen-Nürnberg, Staudtstrasse 7/B2, 91058 Erlangen, Germany} 
\author{Golnoush Shafiee}
\affiliation{Max Planck Institute for the Science of Light, Staudtstrasse 2, 91058 Erlangen, Germany}
\affiliation{Department of Physics, Friedrich-Alexander-Universität Erlangen-Nürnberg, Staudtstrasse 7/B2, 91058 Erlangen, Germany} 
\author{Kaisa Laiho}
\affiliation{German Aerospace Center (DLR e.V.), Institute of Quantum Technologies, Wihelm-Runge-Str. 10, 89081 Ulm, Germany}
\author{\\Dmitry V. Strekalov}
\affiliation{Max Planck Institute for the Science of Light, Staudtstrasse 2, 91058 Erlangen, Germany}
\author{Gerd Leuchs}
\affiliation{Max Planck Institute for the Science of Light, Staudtstrasse 2, 91058 Erlangen, Germany}
\affiliation{Department of Physics, Friedrich-Alexander-Universität Erlangen-Nürnberg, Staudtstrasse 7/B2, 91058 Erlangen, Germany} 
\author{Christoph Marquardt}
\affiliation{Department of Physics, Friedrich-Alexander-Universität Erlangen-Nürnberg, Staudtstrasse 7/B2, 91058 Erlangen, Germany}
\affiliation{Max Planck Institute for the Science of Light, Staudtstrasse 2, 91058 Erlangen, Germany}

\date{\today}

\begin{abstract}
Crystalline Whispering Gallery Mode Resonators (WGMRs) have been shown to facilitate versatile sources of quantum states that can efficiently interact with atomic systems. These features make WGMRs 
an efficient platform for quantum information processing. Here, 
we experimentally show that it is possible to generate polarization entanglement from WGMRs by using an interferometric scheme. Our scheme gives us the flexibility to control the phase of the generated entangled state by changing the relative phase of the interferometer. 
The S value of the Clauser-Horne-Shimony-Holt’s inequality in the system is \(2.45\pm0.07\), which violates the inequality by more than 6 standard deviations.

\end{abstract}

\maketitle


\section{Introduction}

Entanglement plays a pivotal role for several quantum technologies in the field of quantum communication and information.
For these applications miniaturized plug \& play photonic sources are desired. Additionally, the chosen
quantum emitter has to be compatible with the quantum hardware such that the light source can drive the latter.
For example, efficient distribution and sharing of entanglement between different nodes is important for the realization of quantum communication networking infrastructure. In many proposals, the nodes of such a network would consist of atomic or atom-like systems~\cite{sangouard2011quantum}. The optical transitions of those systems 
typically exhibit narrow bandwidths in the range from 10 to 100 MHz. An efficient interface between different nodes therefore requires sources of entangled states with the optical bandwidths on the same order.

Often, one generates entangled states with non-linear processes such as spontaneous parametric downconversion (SPDC) in second-order non-linear materials~\cite{kwiat1995new, rarity1990experimental, kim2006phase, zhu2012direct} or spontaneous four-wave mixing in materials exhibiting third-order non-linearities~\cite{takesue2004generation}. To realize the entangled states, one superimposes the outputs of two independent processes which are highly indistinguishable. 
The two typical configurations are using two nonlinear crystals whose optical axes are perpendicular~\cite{kwiat1995new} or a crystal with an interferometric setup~\cite{rarity1990experimental, kim2006phase}. One drawback of such sources is that the bandwidth is magnitudes wider and not compatible with atomic systems without further effort. In our case, we use a Whispering Gallery Mode Resonator (WGMR) as a source for generating polarization entanglement. Such sources typically have a bandwidth of a few tens of MHz and can interact with narrowband systems without additional filters.


WGMRs are made of highly transparent dielectric materials. When light propagates in such a resonator, it is confined near the rim due to the total internal reflection. There are several interesting advantages to achieving SPDC in WGMRs~\cite{strekalov2016nonlinear}. Firstly, WGMRs have a high Q-factor (\(Q>10^7\)) and small mode volume (less then \(10^6\lambda^3,\) where \(\lambda\) is the optical wavelength inside the resonator), which can increase the efficiency of SPDC. The high Q factor ensures a narrow bandwidth for the whole transparency region of the material. Moreover, the spectrum of WGMRs can be widely tuned by various techniques. Additionally, it is possible to fine tune the bandwidth via the distance of the coupling prisms relative to the WGMR~\cite{fortsch2013versatile}. It has been shown that with these properties one can efficiently tune the generated parametric photons from the same WGMR to narrowband transitions of rubidium and cesium~\cite{schunk2015interfacing}. 
 Those features make WGMRs a potentially promissing source of quantum states of light for developing quantum networks. However, one key aspect of this source is missing: polarization entanglement has not been demonstrated in WGMRs.


In this work, we demonstrate for the first time two-photon polarization entanglement from a WGMR in an interferometric scheme. In this scheme, we leverage the directional degeneracy of the whispering gallery modes. By coupling the pump laser into the equivalent clockwise (CW) and counterclockwise (CCW) modes, we generate the parametric signal and idler photon pairs that populate counterpropagating, but otherwise identical, whispering gallery modes. These photon pairs can be used to yield polarization entanglement. This is achieved by rotating the polarization of the cw-propagating beams and combining the two signal beams as well as the two idler beams on polarizing beamsplitters.
We show that by adjusting the relative phase between the combined parametric beams, we can access various polarization-entangled Bell states. These states can be charaterized by the Clauser-Horne-Shimony-Holt's (CHSH) $S$-parameter~\cite{clauser1969proposed}. The quantum states, for which  this parameter takes the values $S>2$, incorporate two-partite superpositions exhibiting tighter correlations than classical systems can have. For the maximally entangled states this parameter takes the value $S=2\sqrt{2}$. The classical boundary $S\leq 2$ is known as the CHSH inequality~\cite{clauser1969proposed}. We demonstrate a violation of this inequality by more than 6 standard deviations, attesting to the quantum nature of the producing polarization states. Apart from the direct measurement of the CHSH $S$-parameter it is possible to extract it from the visibility of the observed interference fringe. We compare these two $S$-values for the generated state and find them consistent.

\section{Results}

\subsection{Experimental setup}



\begin{figure*}[htbp]
    \includegraphics[width = \textwidth]{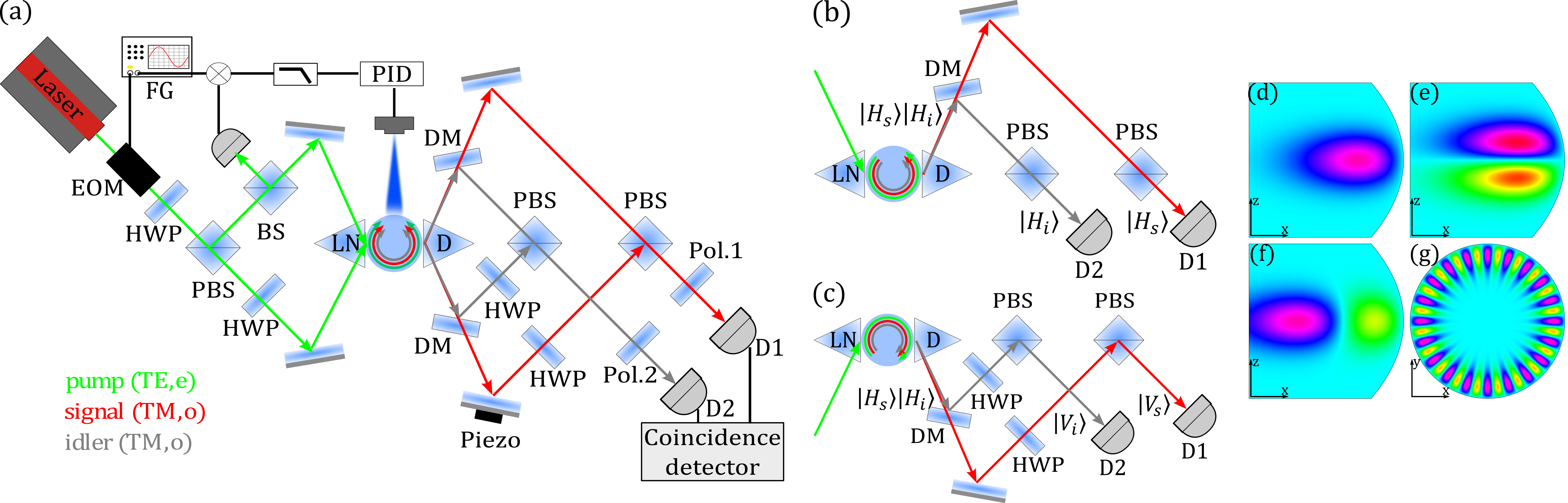}
    \caption{\label{fig:experimental setup}Sketches of the experimental arrangement. (a) is the experimental setup. EOM: electro-optic modulator, FG: function generator, HWP: half-wave plate, PBS: polarizing beamsplitter, BS: non-polarizing beamsplitter, PID: proportional–integral–derivative controller, LN: x-cut LiNbO\(_{3}\) prism, D: diamond prism, DM: dichroic mirror, Piezo: piezoelectric transducer, Pol: polarizer, D: detector. (b) and (c) illustrate the interferomertic arrangement for CCW and CW propagation of the involved beams, respectively. The side views of the electric field distributions are presented for the resonator mode numbers (d) \(L=m\), \(q=1\) (e) \(|L-m|=1\), \(q=1\) and (f) \(L=m\), \(q=2\). The top view of the electric field distribution for a resonator in (g) equals to \(m=20\).}
\end{figure*}


In the following, we explain the experimental setup as shown in Fig.~\ref{fig:experimental setup} (a). We couple a \qty{532}{\nm} continuous-wave laser into the WGMR from the CW and CCW directions. The non-polarizing beamsplitter is used to detect the reflected pump spectrum. The x-cut LiNbO\(_{3}\) prism is used to couple the pump light into the WGMR while minimizing the parametric photons loss due to the parasitic out-coupling~\cite{sedlmeir2017polarization}. The diamond prism is used to couple the signals and idlers out of the WGMR. Note that because of the wavelength difference, the pump loss due to evanescent coupling through this prism is strongly suppressed when the signal and idler out-coupling are optimized. In this way, using two separate prisms to couple the three involved light fields allows us to selectively optimize the coupling rates. The optic elements located on the right side of the WGMR are used to create polarization-entangled states and examine their quality. The WGMR is coarsely temperature stabilized using a Peltier element and a temperature controller. In addition, we implement a fast temperature control technique by shining a blue light on top of the WGMR~\cite{shafiee2020nonlinear}.



Modes of this resonator can be identified with three numbers $L$,$m$ and $q$~\cite{breunig2013whispering}. The first two numbers represent the photon's orbital momentum and its z-projection, respectively, while the radial mode number q is equal to the number of the optical field anti-nodes encountered in the radial direction (Fig.~\ref{fig:experimental setup} (d)-(f)). The type-I non-critical phase matching is achieved by controlling the temperature of the resonator, which fulfills the photon energy conservation between the pump ($p$), signal ($s$) and idler ($i$) modes:~\(\omega_{p} = \omega_{s} + \omega_{i}\)~\cite{boyd2008nonlinear}. The phase matching between a combination of three modes, can be symbolically written as \((L_{p}, m_{p}, q_{p}) = (L_{s}, m_{s}, q_{s}) + (L_{i}, m_{i}, q_{i})\). The selection rules associated with this phase matching are discussed in Ref.~\cite{fortsch2015highly}. The fundamental property of the whispering gallery mode phase matching leveraged in our approach is that if \((L_{p}, m_{p}, q_{p}) = (L_{s}, m_{s}, q_{s}) + (L_{i}, m_{i}, q_{i})\) is achieved, then \((L_{p}, -m_{p}, q_{p}) = (L_{s}, -m_{s}, q_{s}) + (L_{i}, -m_{i}, q_{i})\) is granted for the same set of mode frequencies. Furthermore, the CW and CCW modes differ only in the sign of their $m$ number and have identical spatial profiles for both the internal and evanescent fields. As a result, the parametric photons emitted in the CW and CCW directions are in principle indistinguishable, and the interferometric generation of the polarization-entangled states is enabled. Any deviations from this can be attributed to technical imperfection in the setup.

\begin{figure}[htbp]
    \includegraphics[width = \columnwidth]{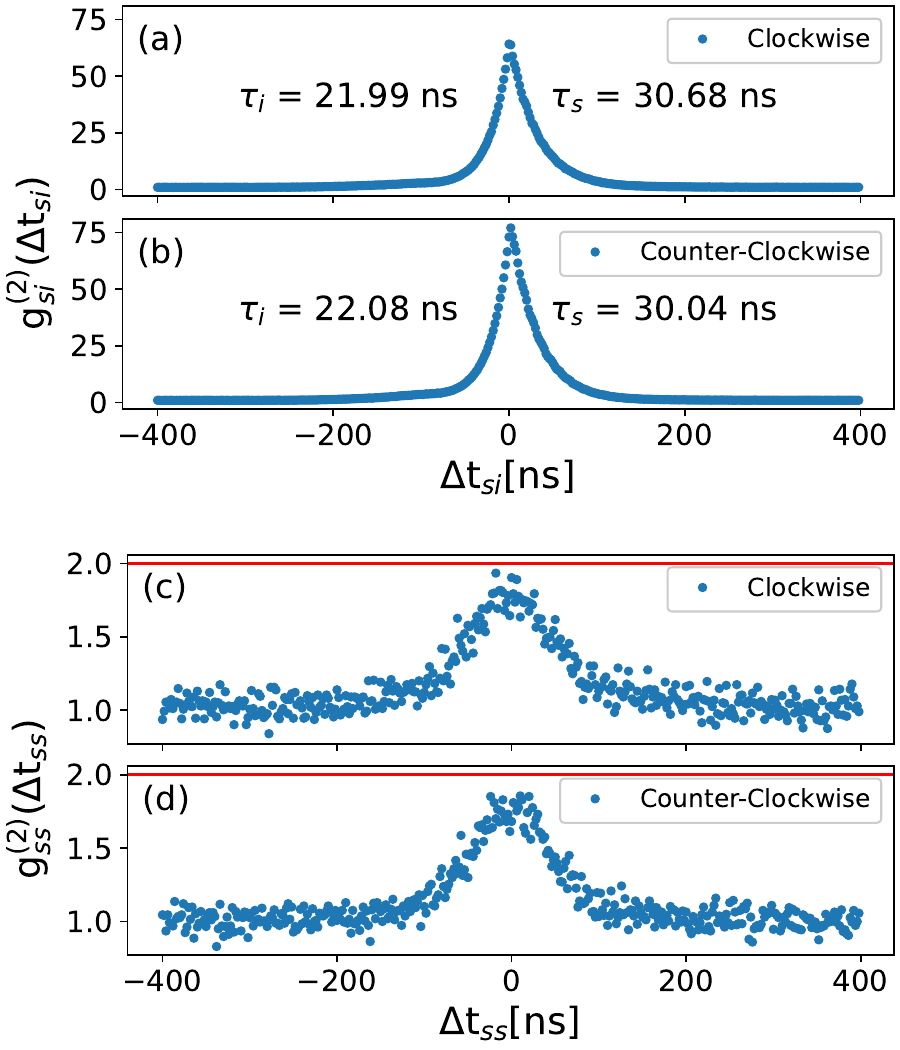}
    \caption{\label{fig:g2data}Conventional characterization of the CW and CCW PDC processes. (a) and (b) represent the cross-correlation functions of the signals and idlers generated from the CW and CCW directions. The leading and trailing time constants \(\tau_{s,i}\)~\cite{ou1999cavity} are calculated by exponential fitting. (c) and (d) show the auto-correlation functions of the signals generated from the CW/CCW direction. The red lines are the theoretical prediction of the peak value for a perfectly single mode quantum state.}    
\end{figure}

Since the most efficient SPDC conversion is achieved for the fundamental modes, when all \(L_{j}=m_{j}\) and \(q_{j}=1\), we adjust the set point temperature (T=\qty{90.21}{\degreeCelsius}) to achieve this type of phase matching. The central signal wavelength is measured at \qty{950}{\nm}, while the central idler wavelength is calculated to be at \qty{1210}{\nm} due to energy conservation. The Q factor of the pump mode is \(1.31\times10^7\) with a bandwidth of \qty{38.7}{\MHz}. The properties of the generated photons are characterized by measuring the cross-correlation and auto-correlation functions in both propagating directions (see Fig.~\ref{fig:g2data}). The bandwidths of the signal and idler are \qty{5.31}{\MHz} and \qty{7.23}{\MHz}, corresponding to Q factors close to \(5.95\times10^7\) and \(3.43\times10^7\), respectively. The pair generation rates normalized to the pump power are \qty{7.29d6}{\per\s\per\mW} in the CW and \qty{1.37d7}{\per\s\per\mW} in the CCW direction. These rates relate to the in-coupled pump and are corrected for the detection efficiency. We also characterize the effective number of modes presented in the CW and CCW beams by investigating the signal-signal autocorrelation function. For a perfectly single-mode SPDC process, one expects a peak value of exactly 2, corresponding to the perfectly thermal photon number statistics~\cite{christ2011probing}.  The measured peak values for the CW and CCW beams are \(1.90\pm0.20\) and \(1.80\pm0.11\) respectively, which show that the states are close to single mode. Note that the single-mode source cannot be frequency-entangled, because it lacks a sufficient number of degrees of freedom among which the entanglement can take place.

\subsection{Experimental results with two-photon interference}
\begin{figure}[htbp]
       
\end{figure}
Fig.~\ref{fig:experimental setup} (b) and (c) illustrate the working principle of how to generate bidirectionally pumped polarization-entanglement from a WGMR. Since we use type-I phase matching and the phase of the generated two-photon state is equal to that of the pump light \cite{graham1968quantum}, the state with a low pump power at the out-coupling prism(D) can be expressed in terms of the pump frequency \(\omega_{p}\), the pump propagation time from the laser to the WGMR \(t_{p}\), and the polarizations of the signal and idler as \(e^{i\omega_{p}t_{p}}|H\rangle_{s}|H\rangle_{i}\). For the CCW beam, after propagating, the state at the PBSs can be expressed as
\begin{eqnarray}
    |\Phi_{CCW}\rangle\propto e^{i(\omega_{p}t_{CCW,p}+\omega_{s}t_{CCW,s}+\omega_{i}t_{CCW,i})}|H\rangle_{s}|H\rangle_{i}
\end{eqnarray}
where \(\omega_{s}\) and \( \omega_{i}\) are the central frequencies of the signal and idler, and \(t_{CCW,s}\) and \(t_{CCW,i}\) are the propagation time of the signal and idler from the out-coupling prism to the PBSs.

For the CW beam, the state has a \(\pi/2\) polarization rotation. The state at the PBSs is
\begin{eqnarray}
    |\Phi_{CW}\rangle\propto e^{i(\omega_{p}t_{CW,p}+\omega_{s}t_{CW,s}+\omega_{i}t_{CW,i})}|V\rangle_{s}|V\rangle_{i}
\end{eqnarray}

Since we perform coincidence measurements, photons from different pairs don't have a correlation and only contribute to the noise. We can represent the state as
\begin{eqnarray}
    |\Phi\rangle = \frac{1}{\sqrt{2}}(|HH\rangle + e^{i\varphi}|VV\rangle) 
        \label{eq:bell-state} \\
    \varphi = \omega_{p}\Delta t_{p}+\omega_{s}\Delta t_{s}+\omega_{i}\Delta t_{i}
        \label{eq:relative phase}
\end{eqnarray}
where \(\Delta t=t_{CW}-t_{CCW}\) and the overall phase is dropped. Note that by moving one of the HWPs from the CW to CCW channels we could produce other types of entangled states, such as \(|\Psi\rangle = (|HV \rangle + e^{i\varphi}|VH \rangle)/\sqrt{2}\).

Our setup is a nonlinear optical interferometer operating at three different wavelengths. A control over the phase \(\varphi\) can be implemented in any part of this interferometer. We choose to do it with a piezo-actuated mirror in the CW signal channel thereby affecting the \(\Delta t_{s}\) in Eq.~\eqref{eq:relative phase}. Besides being actively controlled, the phase \(\varphi\) can drift due to variations in the optical path lengths that need to be stabilized for the duration of the experiment. Furthermore, even for the stable optical path lengths, there is a variation \(\Delta\varphi\) around the fixed value \(\varphi\) arising from the pump, signal and idler frequency fluctuations around their central values:

\begin{eqnarray}
\label{eq:variation phase_laser linewidth}
\Delta\varphi = \Delta\omega_{p}\Delta t_{p} + \Delta\omega_{s}\Delta t_{s} + \Delta\omega_{i}\Delta t_{i}
\end{eqnarray}

Here the pump laser linewidth \(\Delta\omega_{p}\) is very small and may be safely neglected. However the signal and idler resonance widths defining its spectral widths \(\Delta\omega_{s}\) and \(\Delta\omega_{i}\) are not insignificant. This means that the interferometer arm lengths \(L_{CW}\), \(L_{CCW}\) need to be balanced to provide a good overlap of the interfering biphoton wavepackets: \(|L_{s,CCW}-L_{s,CW}|<<c\Delta\omega_{s}\), \(|L_{i,CCW}-L_{i,CW}|<<c\Delta\omega_{i}\). This would be very difficult to achieve with a free-space SPDC which is characterized by very broad phase matching and hence the large signal and idler spectral widths. But with our WGMRs whose signal and idler linewidths are given above it is sufficient to balance the interferometer on lengths that are longer than a few meters.

\begin{figure}[htbp]
 \centering
 \includegraphics[width=\columnwidth]{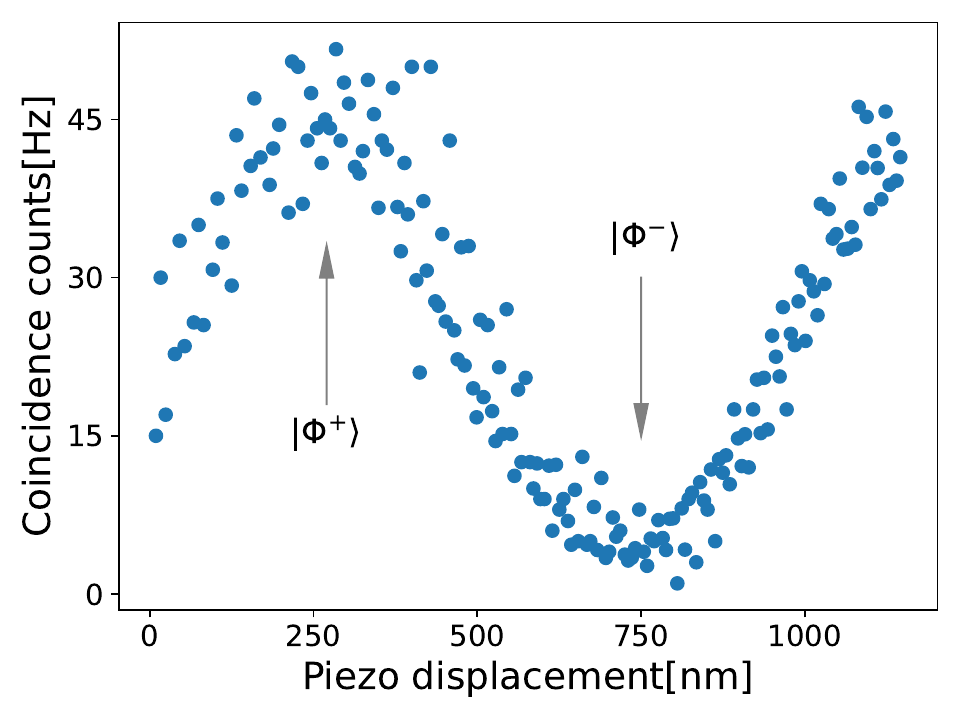}
 \caption{\label{fig:singlet-triplet} Two-photon interference fringe as a function of the phase \(\phi\).} 
\end{figure}

Changing the relative phase \(\varphi\) by applying a voltage to the piezo in the experiment, we can freely change from one Bell state \(| \Phi^- \rangle = (|HH \rangle - |VV \rangle)/\sqrt{2}\) to another \(| \Phi^+ \rangle = (|HH \rangle + |VV \rangle)/\sqrt{2}\).
This is shown in Fig.~\ref{fig:singlet-triplet}. When measured on a diagonal/anti-diagonal polarization basis, the coincidence counts for the state \(| \Phi^+ \rangle\) is expected to reach a maximum while that of the state \(| \Phi^- \rangle\) is expected to reach a minimum. Using the latter setting, we characterized the passive stability of our interferometric setup by tracking the coincidence counts. We found it to be stable on a scale of several minutes, which is sufficient for the following measurements. Details to this can be found in the methods section.

\begin{figure}[htbp]
    \centering
	\includegraphics[width=\columnwidth]{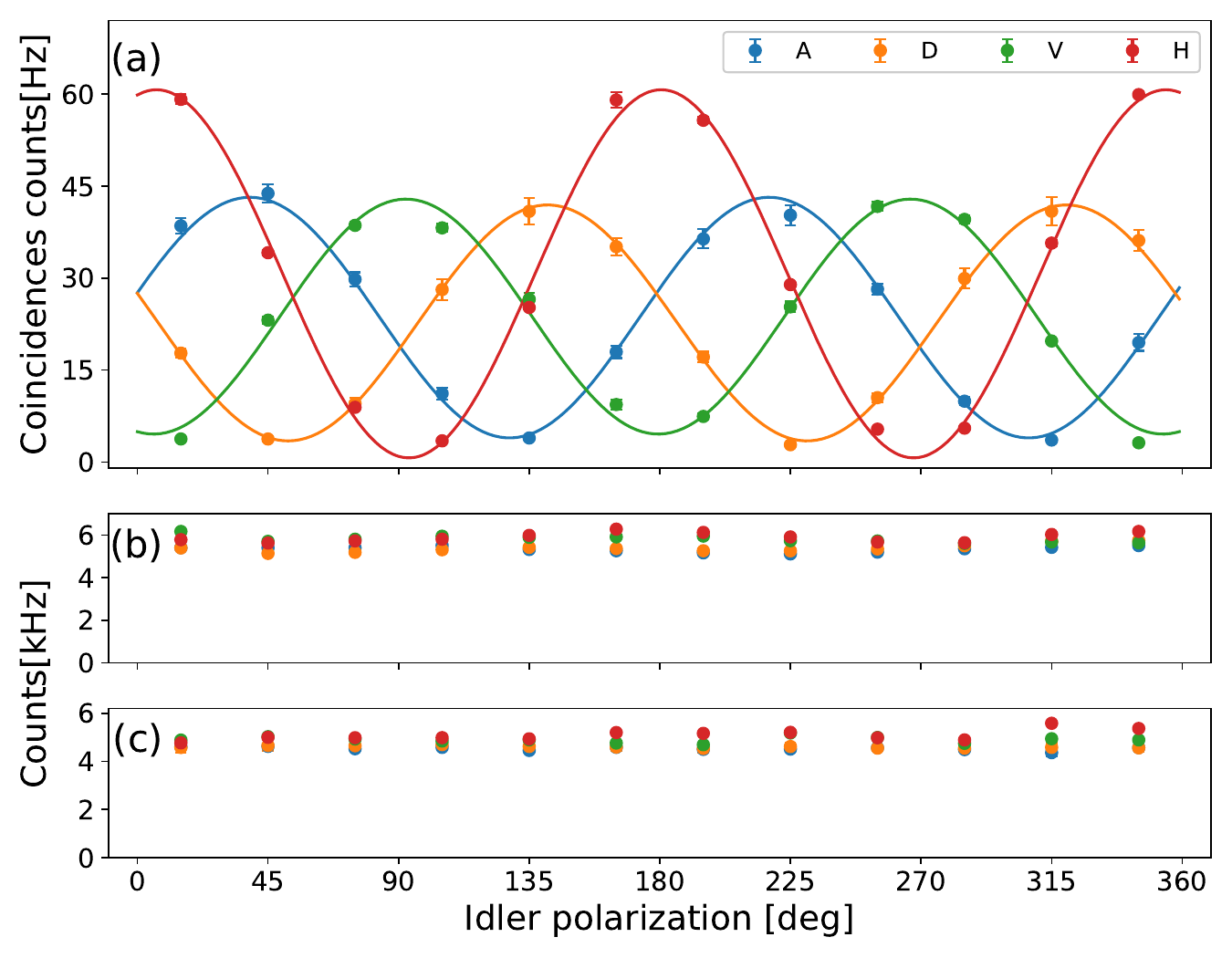}
    \caption{Count rate measurements in terms of rotatitng the polarization in idler beam path. (a) represents the coincidence counts in the four different polarizations H/V/D/A selected in the signal arm. (b) and (c) are the idler and signal counts rate in the 4 different polarization bases.}
    \label{fig:4 sinewave}
\end{figure}

We further demonstrated the non-local two-photon interference through changing the polarization projections. The entangled state we used in the following measurements is \(| \Phi^- \rangle\). The data shown in Fig.~\ref{fig:4 sinewave} are measured with the polarizer in the signal arm set to project the polarization onto the directions \(\theta_s=0^\circ/90^\circ/45^\circ/135^\circ\) (further denoted as the H/V/D/A basis), while changing the projection angle \(\theta_i\) the polarizer in the idler arm. From Eq.~\eqref{eq:bell-state} it is easy to find that the normalized coincidence counts \( \rho \) should fit the prediction
\begin{eqnarray}
\label{eq:sine square}
\rho = \cos^2(\theta_{i} + \theta_{s})
\end{eqnarray}
We fit the data with Eq.~\eqref{eq:sine square}, resulting in a visibility of 95\%/89\%/89\%/86\% in H/V/D/A basis, respectively. These results exceed the classical limit of 71\%~\cite{rarity1990experimental} thus verifying the generation of Bell states. In Fig.~\ref{fig:4 sinewave}(b-c) we further illustrate the single counting rates for the signal and idler detectors recorded during the coincidence counting. By showing that these rates are nearly constant, we emphasize that the observed pattern is the result of a true two-photon quantum interference effect and not a statistical product of two single counting rates. The reason why the maximum value of the coincidences in the H base is higher than that of in the other bases is the difference of the loss in the CW and CCW directions. In the experiments we aimed at keeping similar single count rates in both directions, so we slightly increased the pump power in the CW direction, which has greater losses than the CCW direction.

\begin{figure}[htbp]
    \centering
    \includegraphics[width=\columnwidth]{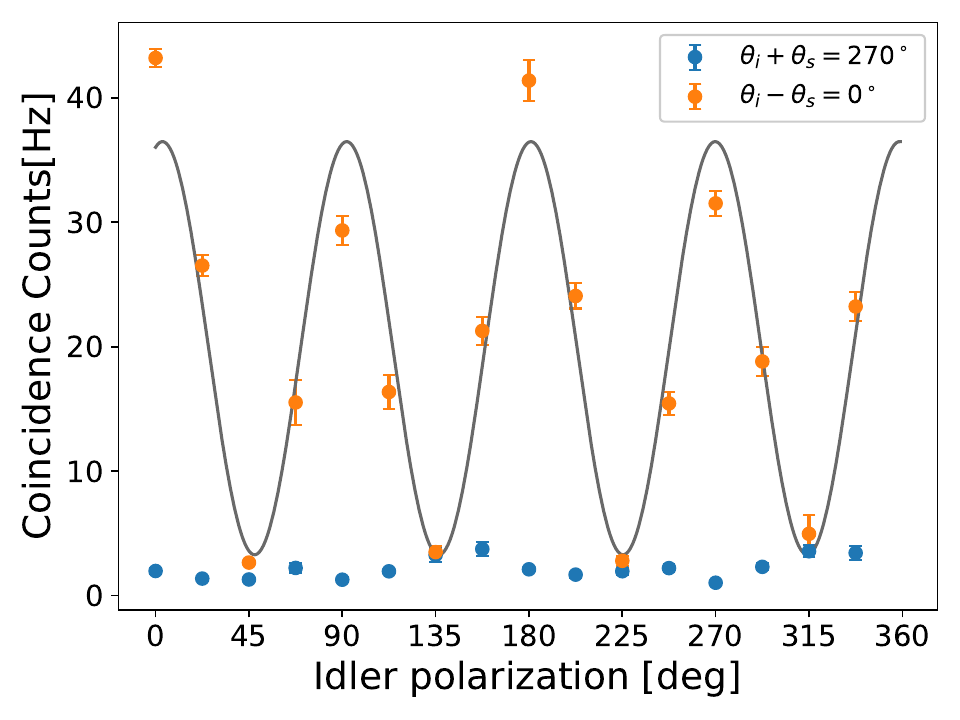}
    \caption{The coincidence counts which are measured when rotating both polarizers at the same time. The gray line is the theoretical prediction of the coincidence counts of \(\theta_{i} - \theta_{s} = 0^\circ\)}
    \label{fig:rotate two HWPs}
\end{figure}

To highlight the nonlocal character of the observed two-photon interference, we rotate both polarizers at the same time in the same and opposite directions. The results shown in Fig.~\ref{fig:rotate two HWPs} fit the prediction in Eq.~\eqref{eq:sine square}.

\begin{figure}[htbp]
    \center
    \includegraphics[width = \columnwidth]{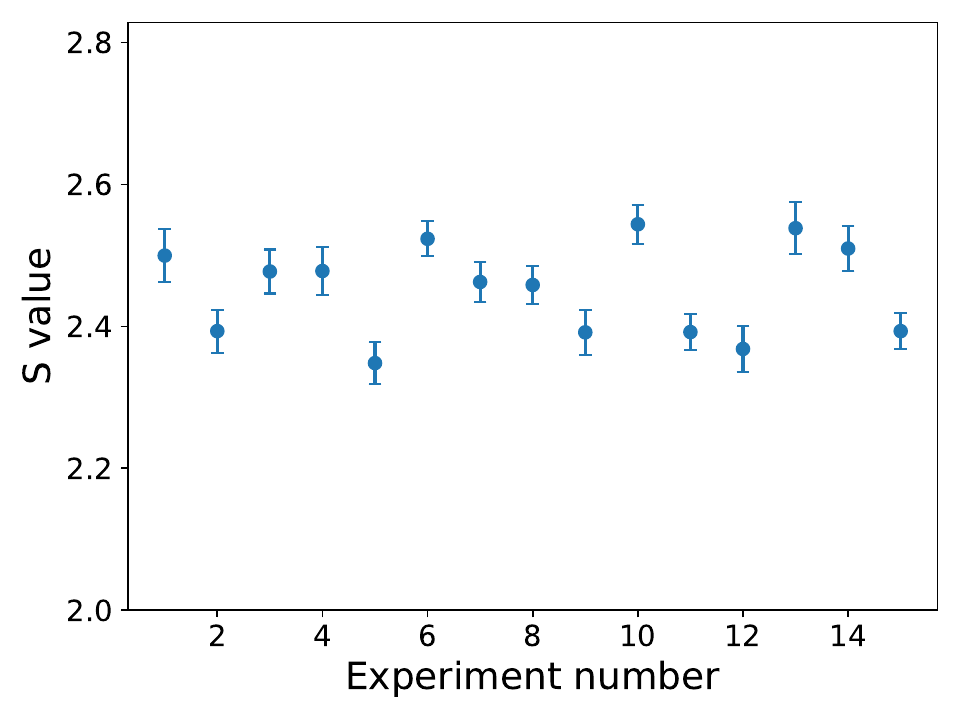}    
    \caption{\label{fig:s_value}The S values extracted from 15 repetitions of the CHSH measurements done within three days. The errors are calculated assuming that the main uncertainties come from the photon counting statistics, and that the statistics are Poissonian.}
\end{figure}

The nonlocal nature of observed two-photon interference allows us to test Bell type inequalities. We calculated the S value of the CHSH inequality \cite{clauser1969proposed} in two ways, arriving at consistent results. On the one hand, we measured the coincidence counts in four polarization combinations, and extracted the S value from the measurements~\cite{kwiat1995new}. The results show \(S = 2.45 \pm 0.07\), which violates the CHSH inequality \(S\le2\) by more than 6 sigmas (Fig.~\ref{fig:s_value}). Additionally, we calculated the S value from the visibilities of the four sinusoidal curves in Fig.~\ref{fig:4 sinewave}(a)~\cite{zhu2012direct, rarity1990experimental}. The result yields a value of \(S = 2.54 \pm 0.06\), which violates the CHSH inequality by more than 8 sigmas.

\section{Discussion}
In conclusion, we have demonstrated the creation of polarization-entangled photons from a WGMR. The fact that the visibilities extracted from the measured coincidence fringes are larger than 85\%, while the single count rates remain unchanged, providing us a genuine sign of the generation of the high-quality entanglement.
The S value of the CHSH inequality is \(S = 2.45 \pm 0.07\), confirming that the generated quantum states are polarization-entangled.

The polarization-entangled photon pairs from the WGMRs can potentially be used for entanglement swapping, in order to distribute entanglement over long distances between the “material” qubits, such as atoms or ions. By design, the signals can meet the frequency and bandwidth requirements of atomic systems for efficient atom-photon interactions, while the wavelength of the idlers is located in the telecom band favorable for long-distance transmission. As a result, long-distance quantum information processing can be realized. Besides, with the same configuration as shown in Fig.~\ref{fig:experimental setup}, it is possible to generate higher-order states, such as 4-photon GHZ states, which makes this type of source interesting for various advanced quantum information applications and protocols. Apart from that, by choosing the coupling regime, one can switch between high photon count rate and long coherence time, making this type of source compatible to applications in two extreme regions. Finally, as the required pump power is low (only a few hundreds nanowatt is needed in this experiment), our source can be enabling for applications with limited power budgets such as space satellite projects.

\section{Methods}
\subsection{Stability of the setup}
\begin{figure}[htbp]
    \center
    \includegraphics[width = \columnwidth]{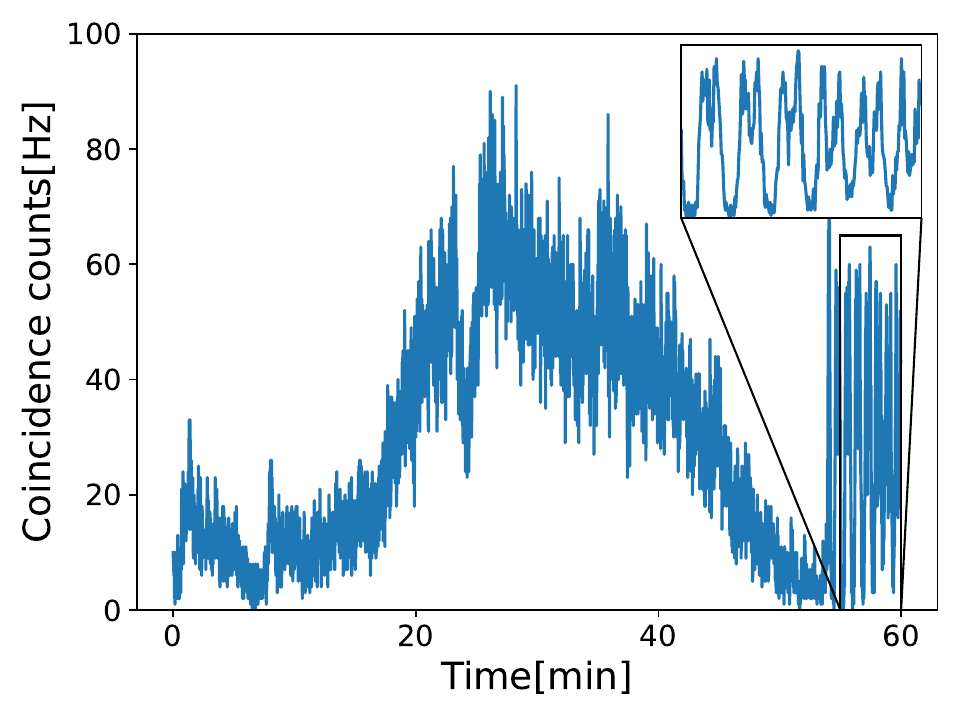} \caption{\label{fig:phase stability}The phase stability is measured by setting both polarizers to \(45^\circ\) and recording the coincidence counts. In the last few minutes we modulated the piezo to know the maximum and minimum coincidence counts, so we can estimate the speed of the phase drift in our setup.}
\end{figure}

We measured the free-running phase drift of our setup to determine how long we can measure without actively stabilizing the interferometer. The result is shown in Fig.~\ref{fig:phase stability}. We found that the drift is slow enough, and we don't need to stabilize the interferometers actively if we finish a single measurement in less than 5 minutes.

\subsection{Coincidence contrasts}
For the measurements in the D/A basis shown in Fig.~\ref{fig:4 sinewave}, we first rotate both polarizers to \(45^\circ/135^\circ\) and apply a voltage on the piezo to minimize the coincidence counts. Since we don't actively stabilize the interferometers, this step is necessary to ensure the state we measured is \(| \Phi^- \rangle\). During the measurements, the voltage applied on the piezo keeps unchanged. After that, we record the coincidence counts of different polarizer settings for \qty{30}{\s} each and then rotate the polarizer in the idler arm by \(30^\circ\). This step is repeated until the idler polarizer is rotated to \(225^\circ/315^\circ\). Then, we minimize the coincidence counts again to compensate for the phase drifts of our setup. After that, we continue to step through the remaining polarizer settings until the polarizer is rotated to \(375^\circ/465^\circ\). This procedure is repeated 10 times in total to gather sufficient statistics for an error estimation.

For the measurements in the H/V basis, the polarizer in the signal arm is rotated to \(0^\circ/90^\circ\) and the polarizer in the idler arm is set to \(0^\circ\). The coincidence counts are recorded for \qty{30}{\s} and repeated 10 times. After that, we rotate the polarizer in the idler arm in steps of \(30^\circ\) and record the coincidence counts until it is rotated to \(330^\circ\).

For the measurements \(\theta_{i}-\theta_{s}=0^\circ/\theta_{i}+\theta_{s}=270^\circ\) shown in Fig.~\ref{fig:rotate two HWPs}, the steps are similar to that of in the D/A basis, but this time we rotate both polarizers simultaneously in the same/opposite direction. Since the oscillation frequency is twice as fast as the frequency in Fig.~\ref{fig:4 sinewave}, we rotate the polarizers by \(22.5^\circ\) instead to catch the feature.


\subsection{S value}
For the measurements of the S value, we first rotate both polarizers to \(45^\circ\) and minimize the coincidence counts with the piezo. After that, we rotate the polarizer in the idler arm to \(67.5^\circ/112.5^\circ/157.5^\circ/202.5^\circ\) in sequence and record the coincidence counts for 30s. We choose these four angles because the greatest violation of the CHSH inequality occurs under these measurements~\cite{clauser1969proposed}. Then, we rotate the polarizer in the signal arm to \(90^\circ\) and the polarizer in the idler arm to \(202.5^\circ/157.5^\circ/112.5^\circ/67.5^\circ\) in sequence and record the coincidence counts. After that, we rotate both polarizer to \(135^\circ\) and minimize the coincidence counts again. Later, we rotate the polarizer in the idler arm to \(157.5^\circ/202.5^\circ/247.5^\circ/292.5^\circ\) in sequence and record the coincidence counts. Then, we rotate the polarizer in the signal arm to \(180^\circ\) and the polarizer in the idler arm to \(292.5^\circ/247.5^\circ/202.5^\circ/157.5^\circ\) in sequence and record the coincidence counts. The measurements are repeated 15 times for the error estimation.

\begin{acknowledgments}
This research was conducted within the scope of the project QuNET, funded by the German Federal Ministry of Education and Research (BMBF) in the context of the federal government's research framework in IT-security "Digital. Secure. Sovereign".
\end{acknowledgments}

\textbf{Author Contributions}
S-H.H, T.D., and G.S. built the experimental apparatus and performed the measurements. S-H.H, T.D., K.L., and D.S. performed the data analysis. G.L. and C.M. supervised the project. All authors contributed to writing the manuscript.


\bibliography{paper_draft}

\end{document}